\documentclass[aps,prd,twocolumn,groupedaddress,nofootinbib]{revtex4}

\usepackage{graphicx}
\usepackage{amsmath,amssymb,amsfonts}
\usepackage[colorlinks,citecolor=blue,linkcolor=blue,urlcolor=blue]{hyperref}
\usepackage[applemac]{inputenc}

\usepackage{physics} 
\usepackage[percent]{overpic} 

\usepackage[english]{babel}

\usepackage[autostyle, english = british]{csquotes}
\MakeOuterQuote{"}

\begin{document}

\title{Neither Presentism nor Eternalism}

\author{* Carlo Rovelli}
\email{rovelli.carlo@gmail.com}
\affiliation{Aix Marseille University, Universit\'e de Toulon, CNRS, CPT, 13288 Marseille, France.\\
Perimeter Institute, 31 Caroline Street North, Waterloo, Ontario, Canada, N2L 2Y5.\\
The Rotman Institute of Philosophy, 1151 Richmond St.~N
London, Ontario, Canada, N6A 5B7.
}

\date{\small\today}

\begin{abstract}
\noindent
Is reality three-dimensional and becoming real (Presentism), or is reality four-dimensional and becoming illusory (Eternalism)?  Both options raise difficulties.  I argue that we do not need to be trapped by this dilemma. There is a third possibility: reality has a more complex temporal structure than either of these two naive options.  Fundamental becoming is real, but local and unoriented.  A notion of present is well defined, but only locally and in the context of approximations.  
\end{abstract}

\maketitle

\section{Introduction}

We usually call "real" what exists in the present, and say that whatever existed in the past (or will exist in the future) is not real now.   \emph{Presentism} is the common sense idea that there is such an objectively real "present" which forms a three-dimensional continuum.  The passage of time, or "becoming" is the continuous transformation, all over the universe, from one objective three-dimensional present instant of time to a new objective three-dimensional instant of time.

The complete empirical success of  special and general  relativity questions presentism, because according to these theories an objective three-dimensional `present' is at best conventional, and at best defined only relatively to a specific motion, hence non objective. (See for instance \cite{Callender2000,Callender2011} and references therein.) It is therefore hard to take it as objectively real. 

The alternative to presentism which is commonly discussed is \emph{Eternalism}. This is the idea that present, past and future event are "equally real". Reality is formed by a four-dimensional continuum. The passage of time, or becoming, is not real, it is in some sense illusory.   

Here I argue that Presentism and Eternalism are both unpalatable, but we are not forced to choose among them;  there is a natural third possibility (see also \cite{Savitt2011,Russell,Harrington2009,Price}), which avoids any tension with relativity without denying becoming.     

The third possibility is the idea that reality has a temporal structure that describes becoming. But this structure is not a simple separation into objective past, present and future.  That is: it makes sense to think that becoming is real, but becoming is different and more complex than a naive oriented one-dimensional succession of instants. 

This possibility has been previously considered ---and I believe it is implicitly assumed by many relativists--- but I have never seen it articulated explicitly and fully. I try to do here.   

There is nothing in relativity which is in contradiction with our experience of time, or that suggests that our experience is "illusory".  What relativity contradicts is the the illegitimate extrapolation of our experience beyond its proper domain. This domain, contrary to our naive intuition, is limited. 

The impact of relativity on our understanding of time has sometimes been taken as a corroboration of classic (pre-relativistic) arguments for the non-reality of becoming. A classic reference for these arguments is Mc Taggart   \cite{McTaggart1908}.  In the last chapter I address  Mc Taggart's argument, and argue ---following the general logic of this paper--- that there is a mistake in it, due to the fact that Mc Taggart's assumptions about the notion of present are excessively and unnecessarily restrictive. Not only there is no contradiction between relativity and becoming, but relativity is itself nothing else than a description of becoming and its structure.  

In this paper I disregard any issue related to the arrow of time ---the difference between the past and the future directions--- or quantum mechanics ---the role of measurement in relation to time---.  I believe that these issues are not truly relevant for the present discussion (on this, see \cite{Zurek2018,Rovelli2016,Rovelli:2015}).  

\section{Presentism and its problems}

Presentism is the idea that there is now a unique real objective three-dimensional present extending all over the universe, formed by the ensemble of the events that are "real now".  As time passes, events in the present become past, while future events becomes present: this is becoming.  

This picture underpins the common way of interpreting non-relativistic spacetime, but is seriously challenged by the empirical success of the two relativity theories. 

The reason, which is well-known, is the following.  In relativistic theories, events that are in the past and in the future of an event $P$ are at time-like distance from $P$ and form a double cone with the vertex on $P$.  The events that are at space-like distance from $P$ are not causally connected to $P$; they form a set ${\cal P}_P$ that can be called the "extended present" of $P$, because any point $P'\in{\cal P}_P$ is simultaneous to $P$ for \emph{some} observer in $P$ according to Einstein's conventional definition of simultaneity.\footnote{$P'$ is said to be simultaneous to $P\!\in\!\gamma$ with respect to a free-falling worldline $\gamma$ if a light ray emitted at $E\!\in\!\gamma$  reaches $P'$  and a light ray emitted at $P'$ reaches $R\!\in\!\gamma$ and the proper time between $E$ and $P$ is  equal to the proper time between $P$ and $R$.}  (See Fig.~1)  ${\cal P}_P$ contains events in the future of one another, in sharp contradiction with the non-relativistic notion of present. 
According to relativity, there is no observer-independent structure of spacetime that permits us to single out a preferred three-dimensional "true present" within this extended present. 

If we demand the `present' at an event $P$ to be: (i) a three-dimensional space-like continuum, (ii) depending only on the (causal) structure of spacetime, (iii) detectable with known physics---then it is a fact that there is no "present" in the universe. 

\begin{figure}[t]
	\includegraphics[width = 6cm]{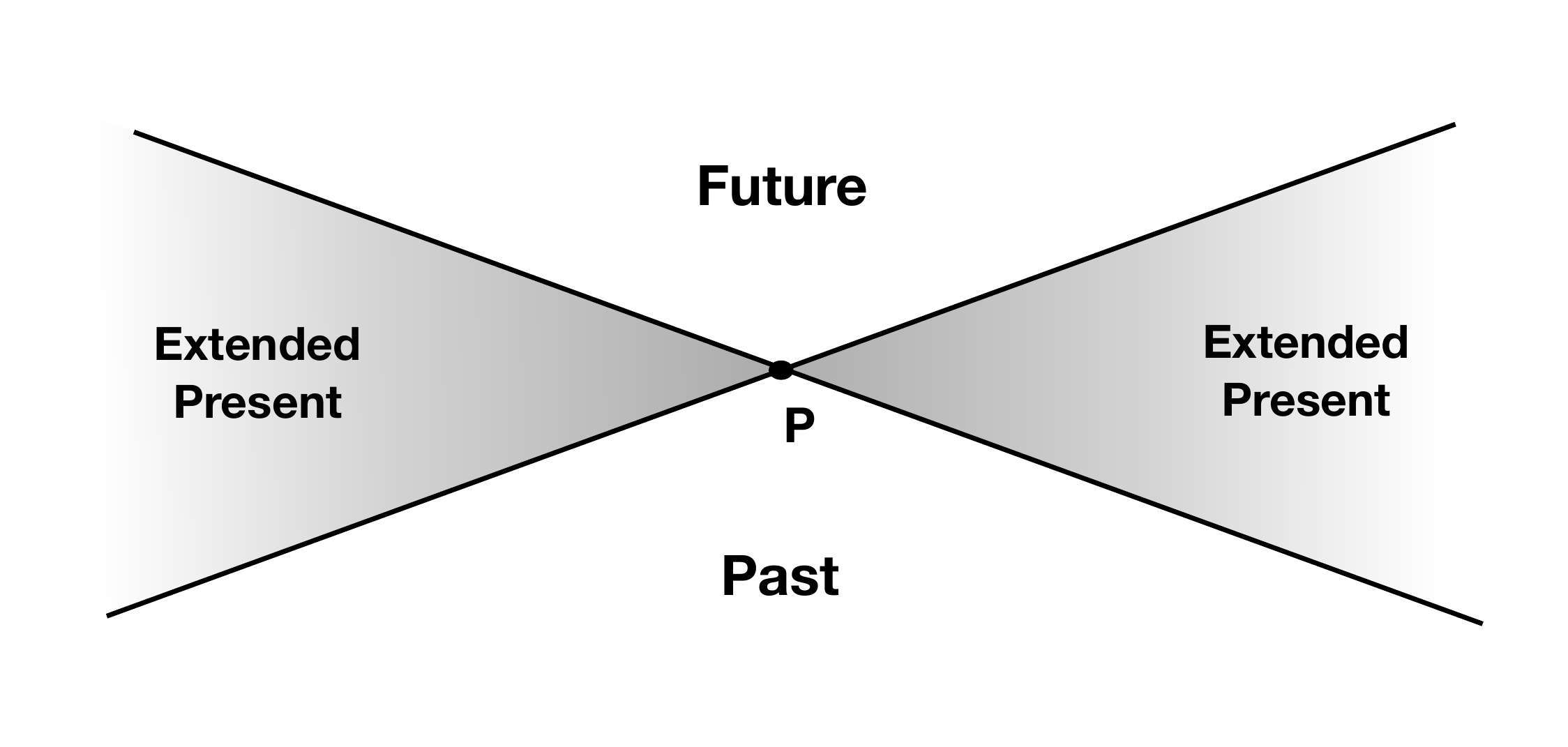}\\
	\caption{Extended present.}
	\label{Extended1}
\end{figure}

\begin{figure}[t]
	{\includegraphics[width = 6cm]{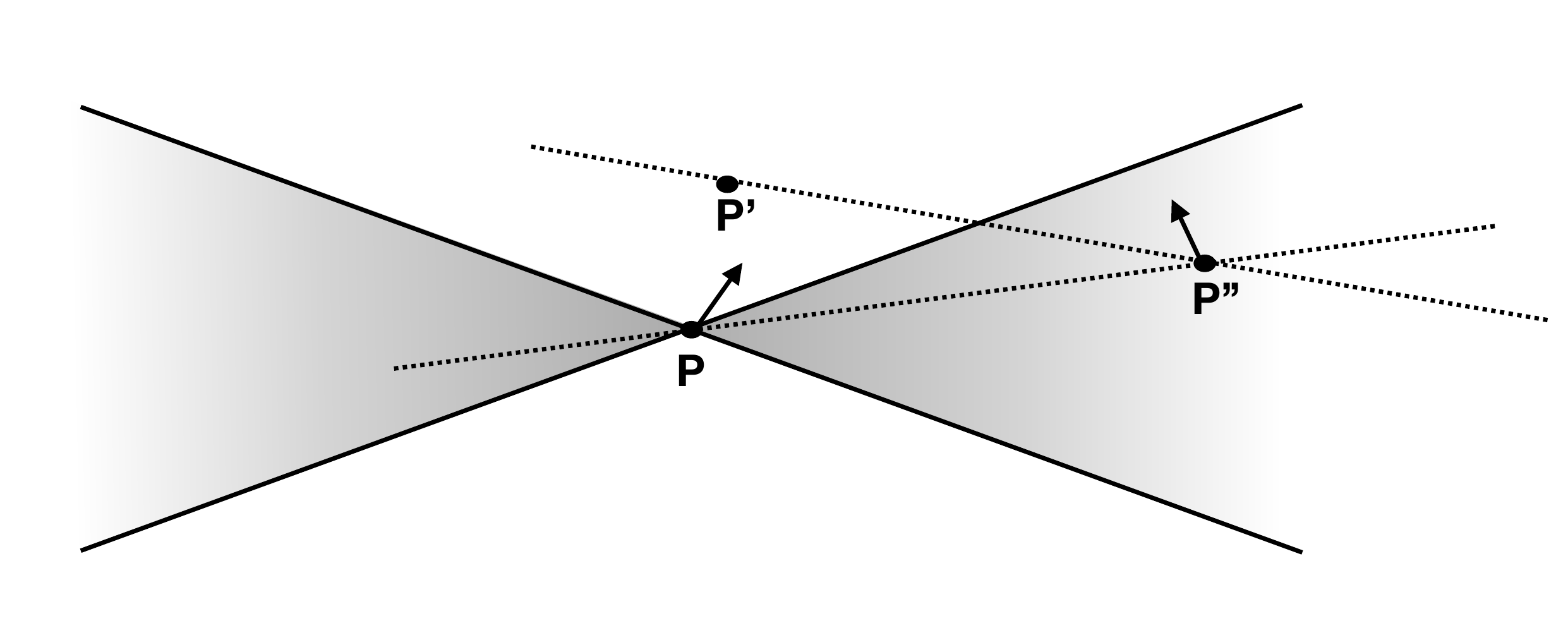}}
	\caption{Putnam argument. The dotted lines indicate simultaneity surfaces relative to distinct observers (arrows); $P$ and $P'$ are "equally real" since they are both simultaneous to $P''$.}
	\label{Extended2}
\end{figure}

We can still define a present if we give up some of these properties.  For instance using preferred matter or a preferred observer. Say we use galaxies to define spatial coordinates and use proper time from the Big Bang along their world-lines as a time coordinate.  But this or similar constructions are conventional and generally fail in the details (two galaxies can meet when their proper time from the big-bang differs, so under this definition a single event could be in the future of itself).
  
Alternatively, we can give up detectability with known physics. That is, we can assume that there is a real present, but is not captured by known physics (see for instance \cite{Craig2001,Crisp2003,Frisch2013,Thyssen2019})\footnote{In EPR-like experiments the effect of a measurement is said to have "instantaneous" effect at space-like distance, conflicting with relativity (see for instance \cite{Callender2007} and references therein). But one cannot even derive the \emph{order} of space-like separated measurements from their quantum correlations. Hence quantum correlations say nothing at all about preferred simultaneity surfaces.}.  This option is considered by some scientists \cite{Smolin:2013ys,Ellis:2013gia,Unger}. I find it unconvincing. The  intuition about the present comes from our experience. Our experience \emph{is}  accounted for by the physics we know. What is the point of trying to salvage an extrapolation of our intuition, if we loose the connection with the reality that generated the intuition?  In other words, this option leads to the bizarre scenario where there is a true present all over the universe, which is not detectable by us, but nevertheless we "know it" from a magical source outside our experience. Which source? 

Barring similar unpalatable steps, Presentism is contradicted by the empirical success of relativity, hence by experience. 

\section{Eternalism and its problems}

Shortly after the formulation of special relativity, Einstein's former math professor Minkowski found an elegant reformulation of special relativity in terms of the four dimensional geometry that we call today Minkowski space.    Einstein at first rejected the idea. ("A pointless mathematical complication".)  But  he soon changed his mind and embraced it full heart, making it the starting point of general relativity, where Minkowski space is understood as the local approximation to a four-dimensional, pseudo-Riemannian manifold, representing physical spacetime. 

The mathematics of Minkowski and general relativity   suggested an alternative to Presentism: the entire four-dimensional spacetime is `equally real now', and becoming is illusory.  This I call here Eternalism.  

A classic presentation of Eternalism and an argument in its favour was given by Putnam in \cite{Putnam1967}. See also \cite{Rietdijk1966}. The core  of Putnam argument is the observation that given any two events $P$ and $P'$ one in the future of  the other, we can always find a third event $P''$ which is simultaneous to $P$ for some observer and  simultaneous to $P'$ for some other observer (Fig.~2).  If we define  "to be real now" as a transitive and observer independent property (as it can be taken to be in non-relativistic physics), it follows that the entire spacetime is "real now". Which is the Eternalist thesis. 

Furthermore, if "becoming" is the continuous transformation from one objective three-dimensional present instant of time to a new objective three-dimensional instant of time, then the absence of an objective present implies that there is no objective becoming.   If there is no objective becoming, then there is something illusory, or at least non fundamental, in the apparent becoming of the world: in the passage of time. 

The problems with Putnam's argument have been repeatedly pointed out.  A particularly clear criticism is in \cite{Stein1968}. See also \cite{Weingard1972,Sklar1985,Shimony1993,Dieks2012,Ben-Yami2015} and especially \cite{Dorato1995}. Putnam misinterprets Einstein's simultaneity and mixes relativistic and non relativistic concepts, making up a mess. In particular, Einstein's simultaneity is not a discovery of a fact of the matter about multiple simultaneity surfaces: it is the discovery that simultaneity has no ontological meaning beyond convention.  This   destroy Presentism, but does not force us into Putnam's Eternalism.   (See also \cite{Saudek2019}.)

A problem with Eternalism as described above\footnote{Different authors have used "eternalism" or "block universe" with different meanings. Sometimes these terms are employed only to indicate any alternative to presentism \cite{Price}. With this I have no objection, or course.} is that it gives a non-dynamical representation of the world.    It fosters an intuition where a four-dimensional universe  "is", instead of "happening".  This is a mischaracterisation of the relativistic theories.  

The Einstein's equations are evolution equations, like any other equations of physics.  There is no reason for not taking them as describing the unfolding of events, coherently with our experience.  The unfolding is not organised by a preferred common time.  But this is not a negation of change: it a description of  change.  

The difficulty with Eternalism (as the idea that past and future event are "real now" as present events) is that it embraces a definition of "to be real now" that clashes manifestly against our common use. It forces us to say that past and future events are "real now", which is nonsense: they are not so, under any reasonable account of the use of "now".   The fact that our intuition cannot be extrapolated does not imply that we cannot use it in its own domain of validity. 

In other words: what is the utility of defining a "real now" in a manner so much in contradiction with its common use?  Relativity questions the role of a three-dimensional objective and universal present, but this does not force us to deny becoming---to think that becoming is not a useful notion to make sense of reality. 
\section{The third option}

When we discover something new about the world, we need to rearrange our vocabulary accordingly, because old words may not match newly discovered facts.  Words denote concepts and concepts follow suite.  

A tribe living in a region where high mountains are in the North may have a concept for "North" which includes the fact that there are high mountains (I happen to be born in such a tribe: the Italians living around Venice). When finding out that there are regions where  North is flat, the concept "North" that included mountains does not work anymore. This does not imply that it must be discarded: it can be corrected, stripping it of unnecessary additions. 

A civilisation that thinks the Earth is flat may have a notion for "up" for which all "up" directions are parallel. On learning that the Earth is round, the notion of "up" must somewhat be revised: "up" in Sydney points to a different star than "up" in London.  This does not mean that "up" is illusory: it means that it works a bit differently than we thought.

Concepts may not survive acquisition of new knowledge untouched.  We have then a choice: keep the old concept strictly as it was, charged with all the implications previously implicitly associated with it.  Or modify it, adapting it to the newly acquired knowledge.  

The first choice forces us to do away with the old concept, because it does not match reality anymore.  We are lead to say  that the notion of "North" is useless when there are no mountains under the Polar Star, or  the notion of "up" is illusory, because it is not agreed upon between Sydney and London.  This choice is silly. 

The good choice is to drop implicit assumptions that are part of old definitions (there are high mountains in the North, all "up" directions are parallel). We can keep using the concepts, adapted to the new knowledge.   We keep their core idea, stripped of the illegitimate assumptions about the world previously packed into their definition. 

We have a similar choice for  "present", "becoming" and  "to be real now".   Reasonably adapting these concepts to the new knowledge pushes us away from Presentism, without forcing us into Eternalism.  

\subsection{The local present}

In common language we say that events happen "now" when, for instance: (a) we watch them happening---including on live TV---, (b) somebody there sees us as we are now, or (c) a third person sees there and us here simultaneously.   These are equivalent definitions of "present" as long as we disregard the fact that light travels at finite speed. 

Concretely, our time resolution is always finite. Without instruments our perception of time can resolve maybe $\sim 0.1 $ seconds.  During this interval, light travels a distance $d\sim 30.000$ kilometres,  a region larger than the Earth.   This means that the three definitions above give the same definition of "now" for all events on Earth, within our resolution of time. This generates our clear intuition of a "present" which is extended in space. 

If we increase our precision in resolving time, the three definitions above agree only over a smaller distance, which defines a bubble of finite radius around us. This can be called the "bubble present". If we measure time intervals with arbitrary precision, the three definitions above agree only over arbitrary small regions.  In the limit of infinite precision, they agree only at a single point.   

If we insist that the "present" is the set of events having all the three properties above with arbitrary precision, the present is  a single spacetime point: the "here now". 

Alternatively, we can relax the definition of "present" in a way that still captures what we indicate in common language, without reducing to a point.  This can be done in different manners. The choice is terminological: it is a matter of convenience, not of ontology.   

Here are some possibilities:
\begin{enumerate} \addtolength{\itemsep}{-2mm}
\item \emph{Einstein's convention}.  Einstein's definition of simultaneity recalled in a note above has the merit of defining time variables with respect to which the Maxwell equations are invariant.  It has the disadvantage of defining a "present" that depends on the world line of an observer. 

\item \emph{Finite bubble present}.  For objects at low relative velocities, and for any given precision $\Delta t$ in the resolution of time intervals, the "bubble present" considered above is defined by a sphere of radius $R= c\Delta t$.  The duration of this present is less that $\Delta t$, hence undetectable.  

\begin{figure}[t]
	\includegraphics[width = 4cm]{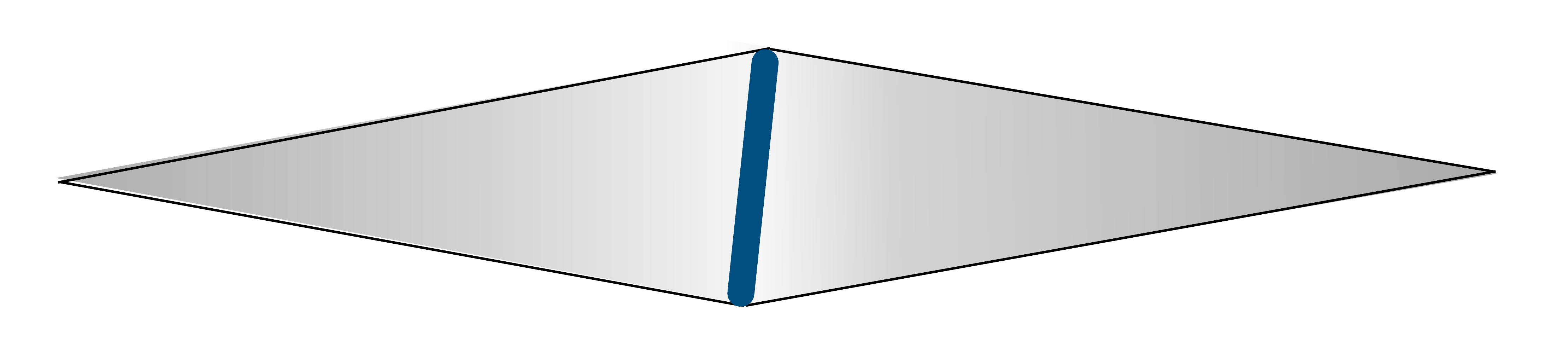}\hspace{.4cm} \raisebox{.05cm}{\includegraphics[width = 4cm]{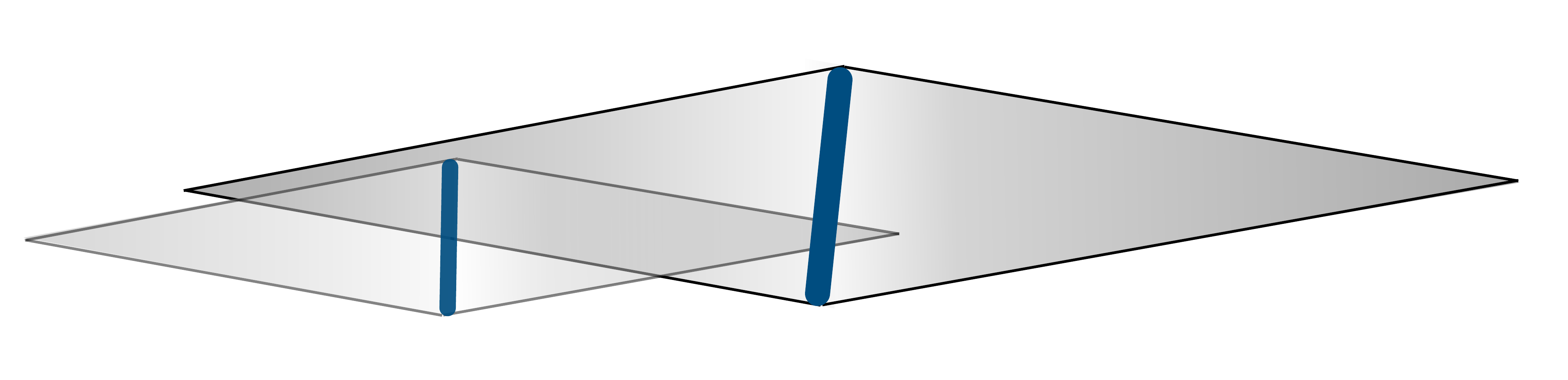}}
	\caption{Left: The diamond-present of an extended event. Right: Two simultaneous extended events: each can send and receive message from the other.}
	\label{Extended3}
\end{figure}

\item \emph{Diamond present}. (I have learned this from \cite{Ben-Yami2019}.)  Contrary to the theoretical physics use, where "events" are defined to be points in spacetime,  in everyday life we commonly use the word "event" to denote happenings extended in time.   A dinner is an event.   Given an event $E$ extended in time (more precisely a compact finite portion of spacetime) its "diamond" region $D_E$ is defined as follows: $P\in D_E$ if and only if there are a $P_-\in E$ which is in the past of $P$ and a $P_+\in E$ which is in the future of $P$.  (See Fig.~3, Left panel)  We can then say that two extended events are simultaneous if each has a point in the diamond of the other. (See Fig.~3, Right panel.) This definition is relative to the two events only and is reflexive.  It captures a common idea of simultaneity.  For instance: "The football match went on during our dinner".

\end{enumerate}

Each one of these definitions matches our common sense use of "present" and "simultaneous" in the everyday contexts.   They show that the common usage of "present" and "simultaneous" is not in contradiction with relativity, provided that it is used within the appropriate approximation and within the appropriate context.  None of them deserves to be charged with ontological weight.  

Relativity is not the discovery of a new ontology of simultaneity: it is the discovery that there is  no fact of the matter, whether two distant punctual events happens at the same time or not. 

\subsection{Becoming}

Physics (if not science in general) is a theory about how things happen.  Its core, since ancient astronomy, Galileo, Kepler and Newton, all the way to quantum field theory and general relativity, is the description of: motion, evolution, change, becoming.  Not "things". The becoming described by physics latches directly to our direct experience of the world as happening.  Thus, becoming is primary both in the phenomenology of our experience and in our physics.  

Notice that what we directly experience is local becoming, not global becoming. That is: we are directly aware of things happening around us, not far away in the universe.  The local becoming that we experience and the becoming well described by Newtonian physics, happen to have a peculiar feature: events can be distinct between past present and future, and labelled by a single time variable $t$ which is tracked faithfully by any good clock, irrespectively from the way the clock moves.  

We are always tempted to extrapolate our experience assuming that what is true locally is true globally. Sometimes this works (the Maxwell equations, found in England, happen to work pretty well in far away galaxies as well), sometimes it doesn't (mountains are not always in the North, the "up" directions are not all parallel and the Maxwell equations are modified in the atomic nuclei.)   In the case of becoming we are tempted to extrapolate local features of becoming to global features: to assume that all events of the universe can be uniquely and objectively separated into past, present and future, and labelled by a single time variable $t$ which is tracked faithfully by any good clock, irrespectively from the way the clock moves.    We have learned that this extrapolation is wrong.  

Hence, if we straight-jacket the notion of becoming into including these features, we are lead to say that there is no becoming in the universe.   

This is a silly choice: analogous to saying that there is no "up" and "down"  because "up" in London is different from "up" in Sydney, or that in Canada there is no true "North" because there are no mountains there (I have heard Italians saying so.)   

The reasonable choice is to recalibrate the notion of becoming, dropping the illegitimate extrapolations implicit in its old conceptualisation.  

This is possible. There is real becoming in the universe.  Things happen. The relativistic equations describe this unfolding of happenings.  Each individual time-like worldline describes a sequence of events, namely a specific unfolding of local becoming.  Distinct local becomings are not independent: they are weaved to one another by the  structure described by the four-dimensional pseudo-riemanniann geometry of general relativity.  The ensemble of all events of the world cannot be objectively arranged into a single simple succession of global instants.  

This impossibility is not the absence of becoming.  It is the fact that becoming is more complex than a naive non-relativistic extrapolation assumes.  The temporal structure of becoming is not the non-relativistic  line with a special point, the "present", but rather the one defined by the causal structure formed by the light cones of a pseudo Riemanniann manifold. 

Relativity does not deny temporality, it shows that it is less trivial than we thought.    
 
The different nows at different locations are not simultaneous: they are independent, and in communication via the causal structure of spacetime.  They are thus partially related, but not fully.  Some "nows" in a distant galaxy are definitely in our past,  some in our future.   But there is a long sequence of distinct "nows" (different moments of time) which are all neither in the past nor the future with respect to the "now here".  This is of course nothing else than Einstein's key discovery: objective simultaneity is meaningless. We can think of reality as a complex web of becoming.\footnote{There is a long tradition of contrary comments by major physicists ("Events do not happen; they are just there, and we come across them", Eddington, 1920. "The objective world simply is, it does not happen", Weyl 1949. "Each observer has his own set of “nows”, and none of these various systems of layers can claim the prerogative of representing the objective lapse of time", G\"odel 1949.  "An observer is merely a world-line, once and for all, on the four dimensional manifold?", Geroch 1984. All quoted in \cite{Dorato1995}.) I disagree with them.  They played a rhetorical role when the new physics needed to break with old habits of thinking, but they are not the best guide for clarity today.  A different case is the  commonly quoted phrase by Einstein: "For us believing physicists, the difference between past, present, and future amounts to an illusion, albeit stubborn" (Einstein to Besso's wife, 21 May 1955), which I believe is misinterpreted,  when taken out of its emotional context (on this, see \cite{Rovelli2018}, Chapter 7).} 

Our common sense intuition about time evolution and becoming is complex and multilayered \cite{Rovelli2018,Cassirer1957}. Different aspects of experiential time depend on different natural structures.  \emph{Some} aspects of our common-sense intuition about time do not carry on to relativistic becoming.  Directionality, for instance, is rooted in the fact that we interact with the world via  macroscopic coarse-grained variables.  It is a property of these variables and it does not belong to the elementary grammar of relativistic becoming.  Hence the fundamental becoming I am referring to is un-oriented  \cite{Price}.   Similarly, our vivid sense of the flow of time is a consequence of the functioning of our brain, rooted in memory and anticipation, and so on \cite{Rovelli2018}.   But the fact that so many aspects of experiential time depend on approximations, and on complex structures, does not alter the fact that what elementary physics describes is happenings, not entities.

\section{What is real now?}

What is real in the universe, then?   The question is ill defined.   Reality has a temporal structure, therefore asking "what is real?" without specifying "when" leads to mixing events that are real now with those that were real in the past but are not  real anymore.    Hence to talk about the reality of something we have to specify a time.   But to specify a time it is not sufficient to specify a number.   We have to locate a region in the rich temporal structure of  the universe.  There are facts that are real now on Earth, facts that were real in the past, or will be real in the future with respect to here-now, and there are also facts that are real on distant galaxy at a time which is neither in our future, nor in our past, but nevertheless they are in the past of one another.   

This may be hard to develop an intuition for, but it is just the way reality is.   Our ancestors had equal difficulty in figuring out how people could live upside down on the other side of the Earth.

The fact that some events can be "real now here" without being "real now" in some other location is no more and no less mysterious than the fact that some events can be "real now" at some time without being "real now" at other times, which so much anguished Mc Taggart  \cite{McTaggart1908}.  Hence relativity does not really add nor subtract much with respect to the pre-relativistic debate on the reality of time. 

\section{The mistake in Mc Taggart's argument}

I close with a simple consideration on the pre-relativistic debate on the reality of becoming, because it sheds light on the mistake leading to the Eternalist perspective. 

Mc Taggart celebrated paper \cite{McTaggart1908} argues against the reality of time by asserting that in order to be defined, the notion of time requires the existence of a series of events (which Mc Taggart calls an A-series) that can be ordered into Past, Present or Future.  This, argues Mc Taggart, leads to contradiction if there is only a single A-Series (because the same event can be both past and future) or to an infinite regress if there is a sequence of A-Series (if we distinguish the A-Series where an event is future from the A-Series where the same event is past).   

There is a mistake in this argument: we \emph{do not} need another A-Series to distinguish the different A-Series (the one where one event is past from the one where the same event is future).  A series of events ordered only according to the notions of "Before" and "After" ---without "now"-- (which Mc Taggart calls a B-series) suffices.  

The reason is that each event of the B-Series determines a different "now", and such a "now" \emph{locally} promotes the B-Series into an A-series (a distinct one for each event).  As soon as this is clear, the main argument of Mc Taggart (it is contradictory for the same event to be both past and future) fails, because the same event is past in one A-Series and future in another, and there is no contradiction in this.  But  Mc Taggart's infinite regression is also blocked, because what distinguished the various A-Series is not another A-Series: it a B-Series.  Hence there is no infinite regression.

Mc Taggart was after a formal definition of a single "now" \emph{not} embedded into a history. He found that this is impossible without infinite regression or contradiction.  This is correct, of course, but why should we expect it to be possible?  Why should we expect that there must be a way to define a preferred "now"  in a atemporal context?    

Mc Taggart was a Hegelian idealist or, better, a "Bradleyan" post-Hegelian idealist, and therefore he believed in the fundamental reality of an atemporal Absolute.  What he was after was the possibility of defining a single "now" from an a-temporal perspective.  He correctly found it impossible, but this does not concern the rest of us, who are not Hegelian idealists and do not need to ground everything on a timeless Absolute. 

Mc Taggart disregards the fact that at the moment at which he writes, or at which we read his paper, we are not outside the universe: we are situated in time.  Hence the notions of Past, Present and Future do not need an extra ingredient to be determined: they are determined by the  temporal location when they are expressed or conceived.  They are indexical.\footnote{To be sure, in a footnote Mc Taggart considers the objection that, in his words, "the present is whatever is simultaneous with the assertion of its presentness, the future whatever is later than the assertion of its futurity, and the past whatever is earlier than the assertion of its pastness". But he rejects this objection on the ground that:  "This theory involves that time exists independently of the A series, and is incompatible with the results we have already reached." The reason he claims that a B-serie is insufficient to provide time ("[some believe that] the real nature of time only contains the distinction of the B series-the distinction of earlier and later") is that he is after a definition of time that can be be given "from outside the universe" and not from within. His argument shows only that there is no time outside the history of the universe.   Which is fine, but is not a denial of the existence of the time that the history of the universe actually describes.}

The time that we call time is the one defined in the universe, not  outside it. 

It \emph{is} possible to regard a temporal series from the exterior.  This is what we do when we say "the story of Anna Karenina", or "The Middle Ages".   The events in this story, or these ages,  form a B-series, seen from an external perspective.  There is nothing wrong in this, because this is precisely a view of a time series from the exterior.  This external view disregards the fact that \emph{at any time of the actual sequence}, the rest of the sequence isn't there: it is the happening itself to be "real" one instant at the time.

This subtle mistake of Mc Taggart is the same mistake as that which lies at the root of Eternalism.   The ensemble of the events of the world is four-dimensional, and we can embrace it within a single image. But this is not a denial of becoming, no more than a single chart of the British royal dynasties is a denial of the fact that events happened in England along the centuries. 

Why does this understanding of the limits of Mc Taggart's argument affect the debate between presentism and eternalism?  Because once Presentism is blocked by relativity, it is a Mc Taggart's like argument that leads to Eternalism.  But this argument is wrong: the fact that there is no preferred objective foliation of four-dimensional spacetime into three dimensional "time instants" is not a denial of becoming: it is only a a denial of a synchronised global becoming.   The "third option" between Presentism and Eternalism is simply what most relativists give for granted: there is no global notion of present, but there is a local becoming, at every point of spacetime.  The "present" is not illusory: it is well defined, but relative to a location: in non relativistic physics, it is relative to a temporal location, in relativistic physics is relative to a spacetime location.  

The four-dimensional spacetime is only a cartography of the relations between these multiple local becomings. 

\centerline{***}

This work was partially supported by the FQXi  Grant  FQXi-RFP-1818. 

\newpage

%

\providecommand{\href}[2]{#2}\begingroup\raggedright\endgroup

\end{document}